\documentclass[8pt,draft,notref,notcite]{article}

\usepackage{color,longtable}
\usepackage{amsmath,amssymb,graphicx}
\usepackage{amssymb,graphicx}

\newcommand{\beq}{\begin{equation}}
\newcommand{\eeq}{\end{equation}}

\newcommand{\be}{\begin{eqnarray}}
\newcommand{\ee}{\end{eqnarray}}

\newcommand{\es}[2]{e^{\,\,\,#1}_{(10)#2}}
\newcommand{\Es}[2]{E^{\,\,\,#1}_{(11)#2}}

\usepackage{amsfonts,amsmath}
\usepackage{latexsym}

\def\p{\partial}

\def\ve{\varepsilon}
\def\cL{{\cal{L}}}

\def\cD{{\cal{D}}}
\def\cE{{\cal{E}}}
\def\cF{{\cal{F}}}

\def\cI{{\cal I}}
\def\cJ{{\cal J}}
\def\cN{{\cal N}}

\def\cO{{\cal O}}
\def\cV{{\cal{V}}}

\def\id{{\mathbb{I}}}
\def\a{{\alpha}}
\def\b{{\beta}}

\def\E{{\rm E}_{10}}

\def\cA{{\cal A}}
\def\cB{{\cal B}}
\def\cL{{\cal L}}
\def\cN{{\cal N}}
\def\cM{{\cal M}}
\def\cV{{\cal V}}

\def\cD{{\cal D}}

\def\cW{{\cal W}}

\def\Es{{\rm E}_{7(7)}}
\def\es{{\mathfrak{e}_{7(7)}}}
\def\Ea{{\rm E}_{8(8)}}
\def\En{{\rm E}_9}
\def\Ez{{\rm E}_{10}}
\def\su{{\mathfrak{su}(8)}}

\def\N8{{\cN\!=\!8}}
\def\ft{{\textstyle{\frac12}}}
\def\dt{{\textstyle{\frac32}}}

\begin{document}

\begin{center}
{\bf \Large $\cN=8$ Supergravity, and beyond $^*$}\\[7mm]
  Hermann Nicolai\\[4mm]
{\sl  Max-Planck-Institut f\"ur Gravitationsphysik\\
     Albert-Einstein-Institut \\
     M\"uhlenberg 1, D-14476 Potsdam, Germany\\[1mm]
 Email: {\tt nicolai@aei.mpg.de}} \\[17mm]
\begin{minipage}{12cm}\footnotesize
\textbf{Abstract:} This contribution gives a panoramic overview of
the development of $\N8$ supergravity and its relation  to other maximally 
 supersymmetric theories over the past 45 years. It also provides a personal 
perspective  on the future role of this theory in attempts at unification.
\end{minipage}
\end{center}

\vspace{10cm}

\noindent
*Invited contribution to the book {\it Half a Century of Supergravity}, eds. A. Ceresole and G. Dall'Agata

\newpage

\section{Introduction}
Maximally extended $\N8$ supergravity \cite{CJ,dWN}  is the most symmetric 
field theoretic extension of Einstein gravity in four space-time dimensions.
It is based on the $CPT$ self-conjugate supermultiplet
\beq\label{multiplet}
{\bf 1} \!\times\! \left[ 2 \right] \,\oplus\,
{\bf 8}  \!\times\! \left[ \dt \right] \,\oplus\,
{\bf 28} \!\times\! \left[ 1 \right] \,\oplus\,
{\bf 56} \!\times\! \left[ \ft \right] \,\oplus\,   {\bf 70} \times \left[ 0 \right] 
\eeq
with particle multiplicities in bold, and spins indicated in brackets.
Modulo gauging, it  is a {\em unique} theory, with a fully worked out Lagrangian
invariant under local $\N8$ supersymmetry, diffeomorphisms and local 
Lorentz transformations.
The fact that supersymmetry alone, unlike all other symmetries
considered in particle physics up to that point, could in this way lead to a 
unique answer immediately caught attention, and was the reason that 
$\N8$ supergravity was quickly embraced as a top candidate for 
the unification of physics in the late 70ies. Even before the actual 
construction of the $\N8$ Lagrangian, noticing that the anticipated 
SO(8) gauge symmetry contains SU(3)$\,\times\,$U(1) as a subgroup, M. Gell-Mann 
tried to identify these gauge symmetries with the color group SU(3)$_c$ and U(1)$_{em}$ of
electromagnetism, the symmetries that are believed to survive to the lowest energies.
However, that attempt fell flat right away,  because the spin-$\frac12$ fermions 
came not only in the form of  color singlets and (anti-)triplets, but also 
as  (anti-)sextets and octets. Independently, 
S.W.~Hawking in his inaugural lecture asked the question whether 
-- and in fact suggested that -- the `end of physics' was in sight \cite{Hawking}.
Although he mentions $\N8$ supergravity only very briefly, it was clear  
that this was the theory he had in mind (as he also confirmed to me 
personally on occasion of my first visit to Cambridge).

However, there were yet more obstacles towards
implementing the theory as a possible model of the real world. 
Not only is its SO(8) gauge group too small to contain the
full Standard Model (SM) gauge group, but like all $\cN\geq 2$ supersymmetric
theories, it does not admit chiral interactions between the fermions
and the vector fields of the supermultiplet (\ref{multiplet}) (the
associated SU(8) $R$-symmetry {\em is} chiral).
Furthermore, after gauging it predicts a hugely 
negative cosmological constant which differs from the measured one 
(now known to be positive) by a factor of $\cO(10^{120})$.  This striking discord
with observation, together with the advent of the heterotic string
which seemed to do much better on all accounts as an ansatz for  physics
beyond the SM, was  one main reason why $\N8$ supergravity was
abandoned as a candidate for unification in the early 80ies  (the other
reason being doubts about whether the theory could be perturbatively 
finite to all orders). Particle physicists' and 
supersymmetry practitioners' interests drifted away from maximal  and non-maximal
extended supersymmetry towards `low energy' ($\cN\!=\!1$) matter-coupled
supergravity models. Since then, the prevalent attitude 
towards maximal supergravity is perhaps  best summarized 
in David Gross'  dictum ``$\N8$ supergravity is not a very interesting theory".  

Is this, then,  the end of the story? Is $\N8$ supergravity a theory that is merely
of academic interest, to be relegated to the dusty storage 
room of failed attempts at unification? In this contribution I will 
retrace some developments since the early 80ies to explain
that there have been significant advances of a less
headline grabbing and more technical nature (for instance 
concerning higher loop computations, or a universal description 
of gauged supergravities by means of the embedding tensor),
as also highlighted in other contributions to this volume. I will
also try to set out a counterpoint to the above view by arguing that
$\N8$ supergravity, given the current state of particle physics, and despite its deficiencies,
may be closer to the truth than is generally thought. 

One of the most important developments in the history of supergravity
was the discovery of a fundamental  link between maximal supergravities 
and the exceptional  symmetries of E-type \cite{CJ}, with E$_{11-D}$
appearing in the reduction of maximal supergravity in eleven dimensions \cite{CJS}
to $D$ space-time dimensions, always in its split real form E$_{n(n)}$
(see \cite{BJ} for a general survey of `hidden symmetries' in
gravity and supergravity). These
symmetries become infinite-dimensional for $D\leq 2$ \cite{Julia}, and indeed
maximal supergravity in $D=2$ was shown in \cite{Nic} to possess a
non-linearly realized $\En \equiv \Ea^{(1)}$ symmetry. The possible relevance of the 
maximal rank hyperbolic Kac-Moody algebra $\Ez$ in the reduction 
to one dimension had already been pointed out in  \cite{Julia}. In yet 
another step (morally corresponding to a reduction to zero dimensions) 
a scheme based on  the non-hyperbolic E$_{11}$ algebra has been 
put forward in \cite{West}. A more concrete realization of $\Ez$ 
with a framework for the emergence of space and time near the cosmological 
singularity was proposed  in \cite{DHN}. This work links the BKL conjecture \cite{BKL},
according to which spatial points become causally decoupled near the singularity,
to an effective reduction to one (time) dimension, and posits that the true symmetry
of a unified theory of quantum gravity and matter is revealed only in a 
`near singularity limit' analogous to the high energy limit of gauge theories 
in particle physics.

The results of \cite{DHN} in particular suggest that we must seek a theory 
{\em beyond} $\N8$ supergravity, in  a  framework beyond space-time 
based quantum field theory, where space-time, and with it concepts such 
as general covariance and gauge invariance would be {\em emergent}. 
This theory (which you may still  call `M theory', if you wish) would necessarily 
differ in substantial ways from string theory in its currently known form, 
but retain some of its core ingredients, such as modular invariance
(see \cite{FGKP} for a beautiful  introduction to some of the pertinent
mathematical concepts). It is my view that this theory should be determined by symmetry 
principles alone, with the maximal rank hyperbolic Kac-Moody algebra 
$E_{10}$ as a key actor, thus highlighting the central role of
hyperbolic Kac-Moody symmetries in future attempts at
unification\footnote{A very similar sentiment was
expressed by Igor Frenkel  in a recent e-mail to me, where he points
out that the great achievement  of the 80ies  was to establish a triangle  
between representations of loop algebras, $2d$ conformal field theory 
and complex analysis, but that in the 21st century there will have to be another
triangle,  now involving hyperbolic Kac-Moody algebras, 
$4d$ quantum gravity and number theory. He concludes by saying ``There is no way
in which the (mathematical) God could miss such an opportunity!".}.
One indication that one should keep looking in this direction is the fascinating link 
beginning to emerge between the bosonic Wheeler-DeWitt operator of maximal
supergravity and the $\Ez$ Casimir operator \cite{KN0}.
The `quantum leap' all the way to $\Ez$ may also be needed for
matching the theory with SM physics.

These ideas also call into question the ultimate relevance of 
space-time supersymmetry in physics. There
is now some evidence that $\Ez$  and its involutory 
`maximal compact' subgroup $K(\Ez)$ `know' everything we have
learnt from maximal supersymmetry, and might therefore
supersede supersymmetry as a guiding principle towards unification.
If a more fundamental theory were to transcend space-time concepts,
supersymmetry in the end might only serve as a  theoretical crutch 
to even more  advanced symmetry concepts which do not admit a space-time 
realization.  And in that case, space-time
supersymmetry (and not just its low energy incarnation) may turn out 
{\em not}  to be relevant to particle physics at  {\em any} scale,
which may explain why we don't see it in Nature!



\section{Construction of the theory}
Originally, simple supergravity was constructed by means of the 
Noether procedure \cite{FFvN,DZ}. This is a step-by-step procedure where one starts 
from the combination of the Einstein-Hilbert action in terms of a vierbein $e_\mu{}^a$
and a spin connection $\omega_{\mu ab}$, and the Rarita-Schwinger
action for a spin-$\frac32$ Majorana fermion $\psi_\mu$,
with an appropriate ansatz for the supersymmetry variations.
One then modifies this `starting action' by adding further terms until full invariance
is achieved, with the final step being the check that the algebra closes
at least on-shell. This  ansatz is greatly facilitated if the iterative procedure
terminates after a finite number of steps. 
It was a major achievement to show that this was also true
for the higher order fermionic terms \cite{FFvN,FreedvN}. After much travail with 
second order formalism, the final $\cN\!=\!1$ supergravity Lagrangian finally takes 
a  rather simple form 
\beq\label{SUGRA}
\cL = \frac14 e e_a{}^\mu e_b{}^\nu R_{\mu\nu}{}^{ab}(\omega) 
\,-\, \frac{i}2 \bar\psi_\mu \gamma^{\mu\nu\rho}  D_\nu (\omega) \psi_\rho 
\eeq
with  the supersymmetry variations 
\beq
\delta_\ve  e_\mu{}^a = i\bar\ve \gamma^a \psi_\mu \quad,\qquad
\delta_\ve \psi_\mu = D_\mu(\omega)  \ve
\eeq
In writing (\ref{SUGRA}) we employ 1.5 order formalism \cite{CW,NT}, which 
means that when checking the superinvariance of (\ref{SUGRA}) one only 
varies the vierbein and the gravitino, but for $\omega_{\mu ab}$ substitutes 
the solution of its equations of motion in the last step.
We then need not worry about the variation of the spin connection because
we can simply discard the term $(\delta \cL/\delta \omega) \delta_\ve \omega$
in (\ref{SUGRA}) (although $\delta_\ve \omega$ in first order
formalism {\em is} explicitly known \cite{DZ}). Elimination of the spin 
connection via its equation of motion yields all the required 
quartic fermionic terms in the Lagrangian and the cubic terms in the 
gravitino variatiion. Furthermore the anti-commutator of two 
supersymmetry variations can be shown to close into a diffeomeorphism, a local
Lorentz transformation and a local (field dependent) supersymmetry 
transformation upon use of the fermionic equations of motion \cite{FreedvN}.
That is, the superalgebra closes {\em on-shell}.

1.5 order formalism is also the appropriate tool for the construction
of higher $\cN$-extended supergravities which always start out like
(\ref{SUGRA}), but now with $\cN$ Majorana gravitinos. However,
Lagrangians become progressively more complicated and more
constrained with increasing $\cN$ (and then also 
require explicit quartic fermionic terms on top of the ones produced by 
the spin connection). The next step was the construction of
Einstein-Maxell ($\cN\!=\!2$) supergravity \cite{FvN}.
$\cN\!=\!4$ supergravity, where for the first time scalar fields appear,
followed soon  \cite{N=4} and provided a first glimpse of a geometrical
structure in the scalar sector, which is governed by a non-linear
$\sigma$-model on the non-compact coset space SU(1,1)/U(1).

However, for $\N8$ supergravity this strategy did not pan out
so straightforwardly, mainly  because  it was not clear what kind
of non-linear structure would appear in the scalar sector. This is why
the theory was first constructed in \cite{CJ}  by torus reduction 
from $D\!=\!11$ supergravity which is polynomial in the three-form
matter field \cite{CJS}. In four space-time dimensions, the 70 real 
scalar fields are described by a complex self-dual 4-form \cite{dWF}
\beq
\phi^{ijkl} = \frac1{24} \ve^{ijklmnpq} \phi_{mnpq}    \qquad \phi^{ijkl} \equiv (\phi_{ijkl})^*
\eeq
which thus transforms in the ${\bf 35}$ representation of the $R$-symmetry SU(8).
In this form the interaction terms involve infinite series expansions in these
fields; a term-by-term Noether type construction is furthermore made more 
difficult by the fact that the scalar fields allow non-linear redefinitions.
The breakthrough came with Cremmer and Julia's discovery
that the theory possesses a non-linearly realized  $\Es$ symmetry \cite{CJ}.
Including 63 scalar gauge degrees of freedom corresponding to the SU(8) $R$ symmetry, 
we have altogether  $133 = 70 + 63$ scalar degrees of freedom, which  coincides 
with  the dimension of $ \Es$. As a consequence the 70 scalar degrees of freedom 
get encoded into  a `56-bein', which in unitary gauge assumes the schematic form
(with $\phi\oplus \bar\phi \in {\mathfrak{e}}_{7(7)} \ominus {\mathfrak{su}}(8)$)
\beq\label{cV}
\cV(x) \,=\, \exp
 \begin{pmatrix} 0 & \phi(x) \\
   \bar\phi(x)  & 0 \\
 \end{pmatrix} \,
  \quad \in \;\; \Es\big/{\rm{SU}}(8)
\eeq
where the 56-by-56 matrix $\cV(x)$ transforms as
\beq\label{cV1}
\cV(x)  \; \rightarrow \; {\rm g}  \cV(x) h^{-1}(x)    \qquad {\rm g}\,\in\, \Es\;,\, h(x) \, \in \, {\rm SU}(8)
\eeq
under rigid $\Es$ and local SU(8) (to be completely precise, the
$R$-symmetry group acts as SU(8) on the fermions, and as
SU(8)$/{\mathbb{Z}}_2$ on the bosons).
In this description the scalar fields are just local coordinates
on the coset manifold $\Es/$SU(8), and as such also subject to non-linear
field redefinitions, {\em alias} coordinate transformations in field space.
The main advantage of (\ref{cV}) is thus that it affords the possibility
to formulate the scalar interactions of the theory in a `coordinate independent' 
manner, in analogy with general relativity. The relevant quantities 
in the Lagrangian  coupling the scalars to the fermions are then 
simply determined from the Maurer-Cartan form
\beq\label{MC}
\cV^{-1}\partial_\mu \cV \,=\, Q_\mu + P_\mu   \qquad  Q_\mu\in\su \;\;,\quad
P_\mu \in \es\ominus\su
\eeq
From this representation we immediately derive the Maurer-Cartan equations 
\beq\label{DP}
\p_\mu Q_\nu - \p_\nu Q_\mu + [Q_\mu , Q_\nu] +  [P_\mu , P_\nu] = 0 \;\;,\quad
D_\mu P_\nu -D_\nu P_\mu = 0
\eeq
where $D_\mu = \p_\mu + [Q_\mu, \cdot\;]$ is the $\su$ covariant derivative.
The latter couples the scalars to the fermions in the form of SU(8) covariant derivatives, 
while $P_\mu$ is the Noether term coupling the scalars to the eight
gravitinos and the 56 spin-$\frac12$ fermions. Remarkably, the $\Es$  Maurer-Cartan
equations were already derived (without being recognized as such)
from the imposition of supersymmetry in B. de Wit's earlier paper \cite{dW}, 
cf. eqn. (4.17) there\footnote{Where  the quantities $Q_\mu$ and $P_\mu$  
 are designated as $\cB_\mu$ and $\cA_\mu$, as in \cite{dWN}.}.

On the vector fields $\Es$ acts as a symmetry generalizing electromagnetic duality,
rotating electric into magnetic vector fields. Because there are only 28 vector fields
to start with in (\ref{multiplet}), these must be supplemented by 28 magnetic 
duals to make up a 56-plet $\cA_\mu^\cM = \big(A_\mu^{IJ}, \tilde{A}_\mu^{IJ}\big)$ transforming 
in the fundamental representation of $\Es$, which puts the theory on-shell. 
The vector equations of motion can then be rephrased in terms of a 
`twisted self-duality constraint', the bosonic part of which takes the form \cite{CJ}
\beq\label{TD}
\cF_{\mu\nu}^\cM = \frac12 \ve_{\mu\nu\rho\sigma}  {\cJ^\cM}_\cN \cF^{\rho\sigma \cN} \;\;
\qquad (\cF_{\mu\nu} \equiv \partial_\mu \cA^\cM_\nu - \partial_\nu \cA_\mu^\cM )
\eeq
where the `complex structure' $\cJ^\cM{}_\cN$ (obeying $\cJ^2 = -{\bf 1}$) involves the symplectic metric
$\Omega_{\cM\cN}$ and the scalar fields via the 56-bein (\ref{cV}).  Alternatively, this equation can be stated 
with `flat' SU(8)  indices, see eqn.(2.13)  in \cite{dWN} where also the  fermionic terms are included.
Their presence is one of the reasons why the largest possible duality 
group $Sp(56)$ \cite{GZ} is reduced to $\Es$.

The only deformation admitted by the $\N8$ Lagrangian is the introduction
of minimal couplings of 28 vector fields (a procedure generally referred to as `gauging').
However, the construction of gauged maximal supergravity took some more time,
because the non-linearities of the scalar sector at first seemed prohibitive.
This required new techniques, in particular maintaining SU(8) in (\ref{cV1}) as a local 
symmetry \cite{dWN}.  Originally, it was thought that the `obvious' gauging with 
gauge group SO(8) and the 28 `electric' vectors from (\ref{multiplet}) is the only one. So the 
discovery of other gaugings with {\em non-compact} gauge groups SO($p,8-p$) (which nevertheless 
maintain positivity of the kinetic terms of the vector fields, thanks to the scalar-vector couplings), 
and yet more with non-semi-simple gauge groups came as quite a surprise \cite{Hull}. 

The proliferation of gaugings at first seems to be in conflict with the 
supposed uniqueness of $\N8$ supergravity. However, we now know that all gaugings 
admit a unified description in terms of the embedding tensor formalism \cite{NS,NS1}
(which also works in other dimensions \cite{IT}).
Namely, for any gauging one replaces the partial derivative in (\ref{MC})
by a gauge covariant derivative that takes the schematic form
(with gauge coupling $g$)
\beq
\partial_\mu \;\; \rightarrow \;\; \cD_\mu = \partial_\mu + g\,\Theta_\cM{}^\cA   \cA^\cM _\mu X_\cA
\eeq
where $\Theta_\cM{}^\cA$ is the embedding tensor  coupling the 56 electric {\em and} magnetic
vectors $\cA_\mu^\cM$ to the 133 generators $X_\cA$ of $\Es$.  {\em A priori} this tensor thus
transforms in the  ${\bf 133} \times {\bf 56} = {\bf 56} \oplus {\bf 912} \oplus {\bf 6480}$ 
of $\Es$. The crucial statement is now that not all these representations are
allowed, but that for a consistent gauging we must satisfy the
linear (supersymmetry) constraint \cite{NS,NS1,dWST}
\beq
\Theta \;\in\; {\bf 912} \qquad \mbox{of $\;\Es$}
\eeq
because otherwise the new supersymmetry variations generated by
the gauge couplings on the bosonic side cannot be cancelled by
corresponding new variations of the fermions
(there is also a quadratic constraint on $\Theta$ to ensure closure of the 
gauge algebra). The essential point is that, provided the constraints
are satisfied the embedding tensor allows to `rotate' the 28-dimensional
gauge group within $\Es$, in such a way that it picks a total of
28 vectors out of the full 56-plet, thus also allowing for mixed
electric-magnetic gaugings. Most remarkably, this
has led to the discovery of a new one-parameter
family of inequivalent SO(8) gauged maximal supergravities \cite{DAIT}. 

In summary, the embedding tensor captures all gaugings and 
thus enables a complete classification of maximal gauged theories 
in four dimensions \cite{dWST}. It works as well in other dimensions
although the classification of gauged supergravities in $D\!=\!3$ still
remains to be completed (where the $\bf 912$ representation of $\Es$ is 
replaced by ${\bf 1} \oplus {\bf 3875}$ of $\Ea$, while for $D\!=\!5$ it is the 
$\bf{351}$ of E$_{6(6)}$), highlighting the central role of the exceptional 
symmetries also for gauged maximal supergravities, even though 
these symmetries are broken by gauging. Because the embedding tensor
representations in various dimensions also appear in level expansions
of $\Ez$, one can even view $\Theta$ as a new `field' and thus 
interpret gauging as being due to a kind of `spontaneous symmetry breaking', 
with the given $\Theta$ as a `vacuum expectation value'.

Gauging modifies and supplements the $\N8$ Lagrangian, and thus
the physics, in important ways. First of all it introduces Yukawa-type couplings 
between the fermions and the scalars at linear order in the gauge coupling $g$,
which can generate mass terms once the scalar fields acquire a 
vacuum expectation value. At second order in the gauge coupling
it produces a scalar field potential which lifts the degeneracy 
of the scalars, whereas in the ungauged theory the scalar vacuum expectation
 $\langle \phi_{ijkl} \rangle$ is classically undetermined and can thus
 be chosen arbitrarily  (this problem reappears in string theory in the guise of
the notorious `moduli problem'  which haunts string phenomenologists
to this day). Generically these potentials are unbounded from below,
but nevertheless admit stable AdS-type extrema if the Breitenlohner-Freedman criterion 
\cite{BF} is satisfied (as is the case for all supersymmetric extrema \cite{GHW}).

For all gaugings of maximal supergravity in four dimensions, the potential 
takes the deceptively simple universal form
\beq
V(\phi) \,=\,  g^2 \left(\frac34 A_1^{ij} A_{1ij} - \frac1{24} A_{2\, jkl}^i A_{2\, i}{}^{jkl}\right)
\eeq
in terms of  two complex SU(8) tensors $A^{ij}_1$ and $A^i_{2\,jkl}$ 
depending on the 56-bein and the embedding tensor, which
together transform in the $\bf{912}$ of $\Es$ \cite{dWN1}. For a 
long time only a handful of stationary points was known, namely all
those preserving a residual SU(3) symmetry \cite{Warner}. Meanwhile,
thanks to enhanced computer power and clever new techniques, it has been shown
that the potential of SO(8) gauged theory has a much richer stationary structure 
with (at least) $\cO(200)$ AdS-type stationary points \cite{Fischbacher}
(see also \cite{DI}), and the
same is true for other gaugings. In all cases, there are only a handful of 
supersymmetric stationary points, whereas all of the non-supersymmetric 
ones are unstable, with one exception. 
Maximal SO(8)$\,\times\,$SO(8) gauged supergravity in three dimensions with 
a similarly compact expression for the scalar potential  \cite{NS1} is in a
league of its own, as it is suspected to possess hundreds of 
thousands of stationary points \cite{Fischbacher1}!


\section{Finiteness: to be or not to be?}
From the very beginning the question of $UV$ finiteness (or not) of
$\N8$ supergravity has been a central issue (see also \cite{BS}
for a more detailed review). Pure Einstein gravity has been
long known to be non-renormalizable from two loops onwards 
\cite{GS,Ven}, with the famous Goroff-Sagnotti counterterm
\beq\label{GS}
\Delta\Gamma_{\rm 2\!-\!loop}  \;= \; 
\frac{209}{2880} \frac{1}{\varepsilon}    \int d^4x \sqrt{-g}
 C_{\mu\nu}{}^{\rho\sigma} C_{\rho\sigma}{}^{\tau\omega} C_{\tau\omega}{}^{\mu\nu}
\eeq
(where $C_{\mu\nu\rho\sigma}$ is the Weyl tensor).
The fact that supersymmetric theories tend to have fewer 
divergences fueled hopes that with more and more supersymmetries
divergences might not only get fewer, but eventually  disappear altogether,
especially for $\N8$ supergravity. Indeed, the two-loop counterterm 
(\ref{GS}) does not admit a  supersymmetrized extension, 
thus ruling out (\ref{GS}) as a divergence for supergravity \cite{Gr,DKS}. However,  already
for three loops there is a counterterm quartic in the Riemann tensor that can be
made supersymmetric \cite{DKS}. This is the  square of Bel-Robinson tensor 
\beq
T_{\mu\nu\a\b} \,=\,C^\lambda{}_\a{}^\rho{}_\mu C_{\lambda\b\rho\nu} \,+\,
^*C^\lambda{}_\a{}^\rho{}_\mu {}^*C_{\lambda\b\rho\nu}
\eeq
which is totally symmetric, traceless and conserved on-shell.
For this reason one can in principle expect all supergravities to diverge 
from three loops onwards. Recall that any $L$-loop counterterm comes 
with $(L+1)$ factors of the Riemann tensor $\propto ({\rm Riemann})^{L+1}$ 
(with a rapidly increasing number of kinematic invariants for increasing $L$).
 
Of course, one would still expect extended supergravities to be better 
behaved than $\cN\!=\!1$ supergravity, for which the 3-loop divergence
certainly appears with a non-vanishing coefficient (which, however, 
has so far not been computed). In the absence of a proper
off-shell formulation which probably does not exist for $\N8$ supergravity,
subsequent investigations were  based on on-shell superspace methods
\cite{BH,Kallosh,HL,HST}~\footnote{Superspace geometry for extended 
 supergravities is nicely reviewed 
  in \cite{Howe} where many further references can be found..}. To be sure,
there are good reasons to harbor doubts about conclusions drawn
on the basis of an on-shell formulation which only knows about
the classical equations of motion, and perhaps a more convincing
argument for the eventual appearance of divergences is simply the expectation
that if no obvious arguments can be found for ruling them out, any counterterms 
that can appear {\em will} appear. For $\N8$ supergravity in particular there was 
considerable discussion about a linearized 3-loop counterterm candidate
(depending on the superfield $\cW_{ijkl}(x,\theta) = \phi_{ijkl}(x) + \cO(\theta)$)
\cite{Kallosh,HL,HST}).
However, that construction was never fully completed as it would
have required defining an integral over a fermionic subspace of the 
full $\N8$ superspace, something that no one knew how to do.

As for $\Es$ invariance of candidate counterterms, the early discussions 
centered on {\em linearized} $\Es$ invariance (that is, invariance under
constant shifts $\delta\phi_{ijkl}(x) =  c_{ijkl}$).
For the on-shell $\cN\!=\!8$ superspace,  the $\Es$  symmetry is not manifest. 
Rather one must solve some of the superspace torsion constraints,  and further 
manipulate the torsion coefficients to derive a new superspace quantity 
which obeys the superspace analog of the integrability constraint (\ref{DP}),
from which the existence of a 56-bein $\cV(x,\theta)$ {\em in superspace} 
can be inferred \cite{BH}. In the end, the superspace formulation of the theory 
thus becomes very similar to the component space formulation! 
With regard to higher order corrections, one must in addition show that 
$\Es$ is a symmetry of perturbatively quantized supergravity, and hence 
a symmetry that must also be respected by possible counterterms,
something that was demonstrated only more recently \cite{BHN}. 
Because $\Es$ is only an on-shell symmetry, \cite{BHN}  resorts 
to the Henneaux-Teitelboim formalism, which takes $\Es$ off-shell but breaks manifest 
Lorentz invariance, as it retains  only the spatial components of the 56 electric and 
magnetic vector fields as independent degrees of freedom \cite{Hillmann} 
(relinquishing  manifest Lorentz is not a critical issue because neither 
Lorentz nor  diffeomorphism invariance can be anomalous in $D\!=\!4$ \cite{AW}).

For maximal  supergravity there is a famous  conjectured relation for the onset of 
$UV$ divergences relating $D_{crit}$ (= dimension where divergences  first appear) 
to $L$,  the number of loops, namely
\beq\label{Dcrit}
D_{crit} = 2 + \frac{14}{L}
\eeq
which predicts the onset of divergences for $L=7$~\footnote{As far as I am aware this
 conjecture is due to Warren Siegel, but I have not been able to confirm this. A more recent (and
 more refined) argument  is given in \cite{Vanhove}.} (with non-integer values
of $D_{crit}$  as they appear in dimensional regularization).
This is indeed a natural conjecture from the point of
view of (on-shell) superspace geometry, with the simple counterterm \cite{Kallosh,HL}
\beq\label{7L}
\Delta\Gamma_{\rm 7\!-\!loop} \,\propto \, \int d^4x \, d^{32}\theta\, \cE
\eeq
where $\cE(x,\theta)$ is the superspace vielbein superdeterminant (Berezinian). 
To see that this is indeed a seven-loop term we need only count dimensions: 
the $\theta$ coordinate has dimension [cm]$^{1/2}$, so we need 16 derivatives 
to saturate the integral over the fermionic coordinates, hence eight Riemann 
tensors (the kinematic structure of the counterterm should follow from expanding out 
(\ref{7L}) to the highest superspace component, but  has never 
been explicitly spelled out). However, it has been shown that  
the $\N8$ superspace volume (\ref{7L}) vanishes on-shell \cite{Bossard}
(and does so also for lower $\cN$).
This casts some doubt on the 7-loop prediction,  
but of course does not at all remove the possibility 
that non-vanishing supersymmetric counterterms will
appear at yet higher orders. Indeed, for $L>7$ many more such invariants 
can be constructed which do not vanish on-shell \cite{Kallosh}, for instance 
by means of the superspace torsion component 
\beq
T^{ij \; k\dot\gamma}_{\alpha\beta} (x,\theta) = \ve_{\alpha\beta} \chi^{ijk\dot\gamma} (x,\theta)
\eeq
which can be contracted in many ways and integrated with $\cE$ over
full superspace to produce yet higher order on-shell counterterm candidates.
However, this proliferation of candidate counterterms  (which gets 
much worse in the  gauged theory \cite{HN}) creates some tension
with the supposed rigidity and uniqueness of the theory. Moreover, it is unlikely
that such counterterms can be made compatible with both non-linear supersymmetry
and non-linear $\Es$ symmetry (see below).

While superspace arguments have thus not led to a definite conclusion
concerning the status of divergences, these developments have now been 
overtaken by stunning computational advances which have
enabled calculations that were unthinkable 40 years ago.
Namely, explicit computations have now become possible thanks to 
completely new techniques developed by Zvi Bern and co-workers,
which in part rely on surprising links between gravity and Yang-Mills theory 
in the form of color-kinematics duality \cite{BCJ}. By means
of these techniques it has now been established that $\cN\geq 5$ supergravities
are finite up to and including five loops, see \cite{Bern3,Bern4,Bern5} 
where many more references can be found.  More  specifically,  
these calculations have shown that up to and including four loops,
the theory does not follow the relation (\ref{Dcrit}) but behaves like 
$\cN=4$ super Yang Mills theory for which $D_{crit} =4 + 6/L$. It is
only at five loops, where color-kinematics duality in its original form
fails, that the divergence appears at $D_{crit} = \frac{24}5$.  
The  vanishing of  chiral anomalies associated with the chiral $R$-symmetry 
groups (S)U($N$) for $\cN\geq 5$ \cite{Marcus} could also be relevant
here, in the same way as the renormalizability of gauge theories 
relies on the absence of gauge anomalies  \cite{CKRT}.
Let us also mention the cancellation of the conformal $c$-anomaly for 
$\cN\geq 5$ supergravities \cite{MN1,Tseytlin}. The significance of this cancellation
is not clear because $\cN\geq 5$ supergravities are not conformal as classical
theories, but the cancellations are sufficiently non-trivial not to be dismissed as 
numerical coincidences, and could be indicative of a  hidden 
superconformal structure \cite{GM}.

It is also important to distinguish between linear and non-linear $\Es$,
as the latter is much more constraining than the former.
The possible relevance of full non-linear $\Es$ symmetry and supersymmetry 
for finiteness has been recently re-emphasized in  \cite{GK,Kallosh0,Kallosh1,Kallosh2} 
and references therein.  One complication is that it is not known how to include 
the dual vectors in an $\N8$ superspace formulation. Maintaining {\em non-linear} 
$\Es$ invariance in the presence of higher order terms requires an infinite 
string of higher order corrections for each `seed counterterm', because (\ref{TD}) 
must be replaced by a {\em deformed twisted self-duality constraint} \cite{BN},
whose compatibility with supersymmetry would require incorporating
the dual vectors into $\N8$ superspace.  Similar comments 
apply  to the counterterm structures derived by analyzing 
`soft limits' of scattering amplitudes \cite{Beisert} (`soft' corresponding
to linearized $\Es$). To work out these higher order modifications and to demonstrate their
compatibility (or not) with non-linear supersymmetry thus looks like a mission 
impossible.  As a consequence, not a single counterterm is 
known that would be compatible with full non-linear 
supersymmetry and non-linear $\Es$ invariance.

While the explicit verification of a  divergence at some higher 
loop order could thus settle the issue of finiteness once and for all, 
$L=7$ unfortunately appears to be out of reach for the time being.
Moreover, it is doubtful whether one can simply extrapolate the 
results concerning $D_{crit}$  obtained so far for $L\leq 5$ from 
non-integer to integer dimensions, where special things  may 
happen \cite{Trnka} ({\em e.g.} in the form of `enhanced cancellations').
In particular, there is no way to dimensionally continue $\Es$ away from $D\!=\!4$
to non-integer dimensions, whence dimensional regularization
cannot `see' the extra constraints imposed by non-linear $\Es$ symmetry.
On the other hand, it is hard to imagine how a proof of all order finiteness 
can be  envisaged with techniques that work only order by order (even though
in \cite{Bern3} large classes of potentially divergent higher loop
diagrams could already be excluded). All this could mean
that we may never know for sure whether $\N8$ supergravity is UV finite or not.

However, in the end, the question of perturbative finiteness of $\N8$ supergravity 
could be rendered moot in a larger framework involving $\Ez$.
Recall that finiteness of superstring theory does not rely on
term-by-term cancellations between bosonic and fermionic loops, but
rather on modular invariance, a symmetry that has no analogue in quantum
field theory. After all, all low energy matter coupled supergravities `derived' from or motivated 
by string theory are badly $UV$ divergent, but no one worries about this,
simply because one generally assumes that string theory will take care
of this apparent non-renormalizability by the time one reaches the Planck scale
(although it does not seem clear how this works precisely).
One may therefore suspect that finiteness of $\N8$ supergravity or some 
extension of it, if true,  relies on a similar mechanism but now involving 
$\Ez$ in a space-time-less context.    

\section{M theory?}
As is well known, maximal supergravities also exist in other dimensions, 
with maximal $D=11$ supergravity \cite{CJS} as the ancestor theory
(for an up-to-date review of Kaluza-Klein supergravity, see \cite{Duff}). 
That theory is certainly not $UV$ finite, but given its central role in
what is called `M theory', the putative non-perturbative unification  of 
superstring theory,  this again  prompts the question  how to embed it 
into another, yet bigger and $UV$ complete theory. String theory, being
perturbative  in $D\!=\! 10$, cannot accommodate 
$D\!=\!11$ supergravity in a straightforward manner; the maximal gauged 
theories are likewise often not embeddable in
$D=10$ supergravities. Rather one must appeal to a web of conjectured  dualities, and more 
specifically, to Witten's famous relation $R\propto g_s^{2/3}$ \cite{W}, according to 
which an 11th dimension `opens up' in the limit of large string coupling. But 
the key question remains what this theory precisely is.

There can be no doubt that the best candidate for a 
non-perturbative formulation, at least in the realm of strings and branes,  
is the maximal $D\!=\!11$ supermembrane \cite{BST,BST1}.
For one thing, it has the extra `strong coupling dimension' already
built in from the very start. Furthermore, it can accommodate $D\!=\!11$
supergravity as its massless sector, at least in principle, and with the correct
SO(9) helicity assignments.  But supermembrane theory is much
harder than string theory! While the gravitational part of the superstring 
path integral can be reduced to a {\em finite}-dimensional (albeit still extremely tricky)
integral over (super-)moduli space, and the path integral involving the target space 
coordinates is Gaussian (at least for simple backgrounds), there are no such 
simplifications for the supermembrane. There is no gauge which linearizes the
equations of motion, and fixing all gauges of the gravitational  path integral
in a Polyakov-type approach leaves a nasty non-Gaussian functional integral.
This is why progress with (quantum) membrane theory has been very slow over the past
decades. A major advance was the work of \cite{G,Hoppe} on bosonic membrane
theory, where it was shown that in a Hamiltonian formulation, this theory can be 
reformulated as the $N\rightarrow\infty$ limit of a quantum mechanical 
SU($N$) matrix model.  Supermembrane theory can be similarly reformulated 
as the $N\rightarrow \infty$ limit of a maximally  {\em supersymmetric} SU($N$) matrix model, 
which coincides with the dimensional reduction of maximal $\cN\!=\! 4$ 
super-Yang-Mills theory from ten to one (time) dimension \cite{dWHN}.
Although at that time the term `M theory' had not even been coined yet, this 
work clearly aimed at developing a computable approximation scheme for
a non-perturbative framework beyond string theory, also encompassing 
$D\!=\!11$ (quantum) supergravity. Subsequently it was shown that the supermembrane 
Hamiltonian, unlike the bosonic membrane Hamiltonian,  has 
a continuous spectrum \cite{dWLN} (see also \cite{Smilga}), and, following
unpublished work of J. Goldstone,  target space Lorentz symmetry generators
can be defined for the supersymmetric theory \cite{dWMN} such that classical Lorentz
symmetry, which is broken for finite $N$, is recovered in the $N\rightarrow \infty$ limit
(the corresponding quantum calculation remains a wide open challenge).
Finally, candidate expressions for light-cone supermembrane vertex operators
governing the emission of massless states from the supermembrane
were derived in \cite{DNP}, from which
also candidate expressions for matrix vertex operators can be deduced.

Supersymmetric SU($\infty$) matrix quantum mechanics was taken up again several years later with 
a different interpretation in terms of $D0$-branes~\cite{BFSS}~\footnote{
 After which the supersymmetric SU($\infty)$ matrix model came to be called the `BFSS model', 
  in a classic confirmation of  Arnold's principle.}. Amongst other
things, following \cite{dWLN}, this work led to the important insight that the quantum supermembrane 
is not a one-particle theory, and that the usual two-step strategy of quantum
field theory, with first quantization as the first step,  and second quantization 
and a multi-particle Fock space as the second step, does not work for the
supermembrane (although it still does seem to work for superstrings in the
context of string field theory). In the limit $N\rightarrow \infty$, the distinction between the 
supermembrane and the $D0$ particle interpretation anyhow becomes 
irrelevant: that limit, if it exists, {\em is} nothing but the quantum supermembrane
in flat target space-time. This is a deeply non-perturbative theory: 
one way to think about it is the following analogy
with QED, which is conventionally formulated in terms $n$-particle states of electrons, 
positrons and photons and their scattering amplitudes. If, however, as a gedanken
calculation and ignoring complications with infinite renormalizations,
one were to try to set up and solve a Schr\"odinger-type functional equation 
for a QED wave functional, an energy eigenfunctional would look nothing like 
a state of definite particle number (in accord with the intuition that physical excitations 
in quantum gauge theories are accompanied by infinite clouds of virtual particles).
Quantum gravity is much more than graviton scattering amplitudes! 

Nevertheless, in spite of some recent progress \cite{Mal} (as well
as  some evidence that the existence of the $N\rightarrow \infty$ limit and thus the
consistency of the quantum membrane require supersymmetry \cite{OL}
and eleven target space-time dimensions \cite{MN2}), it appears that following 
this line of thought has so far not gotten us any closer to `real physics', 
if we stay within the realm of maximally supersymmetric relativistic extended
object theories (strings and branes).  Even if the existence of the $D=11$ quantum 
supermembrane could be established, we still would not know how to relate 
it to SM physics. This may be due  to remaining unsolved challenges, especially with 
supermembrane theory, and the fact that the theory may look completely
different in the non-perturbative regime. However, it seems more likely 
that essential elements are still missing, such as for instance concerning the 
question of  whether and where infinite-dimensional hidden Kac-Moody symmetries
could enter the stage. For starters, this would require bringing some order into 
the zoo of $D$- and $M$-branes, Kaluza-Klein monopoles and other 
supersymmetric extended objects which are a central focus of current research
in string theory \cite{HT}. This clearly suggests that one should look for a 
larger framework beyond the supermembrane encompassing membranes, 
five-branes, {\em etc.} These extended objects should 
couple to 3-form and 6-form fields, as well as infinitely many further fields 
associated with more and more non-trivial Young-tableaux beyond $p$-forms. 
However, a cursory glance at the relevant representations arising in an SL(10)
level decomposition of $\E$ \cite{FN} (where the 3-form and 6-form representations 
appear at the very top of an infinite list) should suffice to put in evidence 
the enormity of this challenge!

\section{The real world}
In approaching the problem of unification one faces a basic dilemma. 
The `Einsteinian' point of view (first expressed by Einstein himself
in 1929 \cite{Einstein})  posits that such a theory 
should not only explain {\em how} Nature works at its most fundamental level, but
also {\em why} Nature is the way it is, {\em and not otherwise.}
When adopting this line of thought, one must consequently look for a more 
or less unique theory. $\N8$ supergravity (or the supermembrane)
therefore seems to be on the right track as far as the general philosophy is concerned.
But to be the absolutely correct theory it would have to hit the nail precisely on its head.
Given the numerous consistency requirements and the rich input
from particle physics and astrophysics data that must be matched with the theory,
getting the right answer on the basis of purely theoretic considerations 
would be like winning a mega-jackpot in the lottery! 
If, on the other hand, the theory deviates in substantial 
ways from observation, mismatches are basically impossible to rectify precisely
because of the (desired) uniqueness and rigidity of the theory which
exclude any kind of adjustments that could eliminate conflict  with observation. 
On the face of it  $\N8$ supergravity  must therefore be discarded 
for the reasons already mentioned in the introduction.

String theory 
has taken the opposite road. This was not by design -- the original paper on
heterotic compactification \cite{CHSW} very clearly
expresses the hope that  there should be an almost unique path
from the heterotic string to SM physics --  but simply due to subsequent 
developments of the theory.  Contrary to initial expectations,
string theory has turned out to provide a very large framework, in which, 
subject to some consistency constraints (such as anomaly cancellations) 
almost any effective $\cN\!=\!1$ supersymmetric low energy theory 
containing the SM can be engineered as the outcome of some 
orbifold or Calabi-Yau compactification, or (intersecting) brane construction.
On top of the desired SM spectrum one invariably ends 
up with numerous new degrees of freedom whose non-observation 
necessitates elaborate and often contrived efforts to explain them away,
or at least lift them above some currently unreachable high energy threshold.
The price one has to pay for the huge flexibility in fitting parts of the theory with observation
is a vast landscape with possibly $10^{272000}$ vacua \cite{WT},
and has in its wake led to the idea that the search for unification must be re-calibrated
towards a multiverse-type interpretation, as first (and to me as a non-believer, most
convincingly) argued in A.N. Schellekens' 1998 inaugural lecture  
\cite{ANS}. It has also led to absurd  contentions that the theory does 
not even need experimental confirmation any more. String theory  
in its currently practised form also fails to explain our non-supersymmetric 
low energy world, in spite of a 40 years' collective intellectual 
effort without precedent in the history of theoretical physics.

Efforts at unification are severely hampered by the fact that 
Nature remains tight-lipped about what comes after the SM of particle physics.
After more than 15 years of operation LHC has not produced
a shred of evidence of new degrees of freedom 
beyond those already contained in the SM spectrum, in particular
confirming the absence of low energy supersymmetry in previous
collider experiments (not to mention failed axion and dark matter
searches over the past 40 years). The RG evolution of the measured
SM couplings shows that they can be safely extrapolated to the Planck scale 
(the fact that the Higgs coupling might dip below zero at some intermediate 
scale $\sim 10^{11}$ GeV can be cured with relatively minor
modifications of the scalar sector). This indicates that the SM could 
survive more or less {\em as is} all the way up to the Planck scale (though
certainly not beyond),
a possibility that attempts at unification must now seriously take 
into consideration\footnote{Superstring phenomenology still seems to be 
 largely predicated on the assumption of $\cN\!=\! 1$ supersymmetry,
 with little attention to non-supersymmetric string vacua, and apparently no place for
 a scenario with only the known SM degrees of freedom up to the Planck scale.}.
This would then require a theory that explains only what we 
see, nothing more and nothing less. In that case I claim that
something close to $\N8$ supergravity could be true, as I will now explain.
If, on the other hand, a future collider were to reveal the existence of 
new fundamental spin-$\frac12$ degrees of freedom, such as,
for instance, a fourth generation of quarks and leptons, or simply
a fourth (sterile) neutrino, or any new spin-$\frac12$ fermion predicted 
by low energy supersymmetry, the ideas sketched below would 
be immediately falsified.

In 1983, M. Gell-Mann pointed out a remarkable agreement 
between the $\N8$ spin-$\frac12$ spectrum and the
SM fermions \cite{GellMann}, an agreement that has become even more 
compelling now after the no-show of  `new physics' at LHC.  Noticing that, 
after the removal of eight Goldstinos from (\ref{multiplet})
one is left  with the right number $48 = 3 \times 16$ 
of spin-$\frac12$ fermions (corresponding to three generations
of quarks and leptons, including right-chiral neutrinos),
he came up with the strange idea that the supergravity SU(3) should
be identified with the diagonal subgroup of SU(3)$_c$ and a hypothetical
family symmetry SU(3)$_f$. Putting the quarks and leptons into the 
appropriate representations of SU(3)$_c$ and SU(3)$_f$ 
he obtained exact agreement! In addition, the electric charge 
assignments almost work, except that the U(1) charges are 
systematically off by a `spurion charge' shift of $\pm \frac16$. So the proposal 
fails only by very little, and in a very systematic fashion. Subsequently,
it was shown that this scheme is {\em dynamically} realized at one 
of the stationary points of the $\N8$ potential \cite{NW}.  At that point, 
however, all attempts to push the agreement further failed.

Nevertheless, given the current state of particle physics with only 48 fundamental 
spin-$\frac12$ fermions, it is worth asking, in a first step, what it would take 
to correct the mismatch of electric charges.  Remarkably, this can be  achieved 
by means of the following simple deformation of the U(1) group by the generator 
\beq\label{cI}
\cI \,:=\, \frac{1}{2} \Big(T \wedge \id \wedge \id + 
\id \wedge \, T \wedge \id + \id \wedge \id \wedge \,T + T \wedge T \wedge T \Big) 
\eeq
acting on the tri-spinor $\chi^{ijk}$ of $\N8$ supergravity \cite{MN3}.
Here, the SO(8) matrix $T$ defined by
\beq
T \,:=\,
\begin{pmatrix}
0 & 1 & 0 & 0 & 0 & 0 & 0 & 0 \\
-1 & 0 & 0 & 0 & 0 & 0 & 0 & 0 \\
0 & 0 & 0 & 1 & 0 & 0 & 0 & 0 \\
0 & 0 & -1 & 0 & 0 & 0 & 0 & 0 \\
0 & 0 & 0 & 0 & 0 & 1 & 0 & 0 \\
0 & 0 & 0 & 0 & -1 & 0 & 0 & 0 \\
0 & 0 & 0 & 0 & 0 & 0 & 0 & 1 \\
0 & 0 & 0 & 0 & 0 & 0 & -1 & 0
\end{pmatrix}
\; .
\eeq
represents the imaginary unit in the SU(3) $\!\times\!$ U(1) breaking, 
which in  turn implies $\cI^2=-\id$ (note that $\cI$ still commutes
with the residual SU(3)$\,\otimes\,$U(1) of supergravity). Because the triple 
wedge product $T \wedge T \wedge T$ is \emph{not} an SU(8) element, 
there is no way of incorporating this  deformation into $\N8$ supergravity. 
The `spurion shift' associated with $\cI$ and needed for matching theory
and observation is simply not compatible with supersymmetry.

But now the remarkable fact is that, at least in the one-dimensional 
reduction, this deformation (\ref{cI}) {\em can} be incorporated into $K(\Ez)$,
an infinite prolongation of the $R$-symmetries of maximal supergravities \cite{KN}
(in fact, $K(\Ez)$ is large enough to accommodate the full  SM 
gauge group, and to `undo' the restriction to the diagonal subgroup
of SU(3)$_f$ and SU(3)$_c$ \cite{MN4}). This could mean that matching the charge 
assignments of the SM fermions with a more fundamental theory may 
require bringing in the full glory of hyperbolic Kac-Moody algebras
and their involutory (`maximal compact') subalgebras!
Then the question is no longer whether $\N8$ supergravity is
the  right theory, but  whether and how $\Ez$ and $K(\Ez)$  
can replace space-time supersymmetry as a guiding principle, and what 
kind of pre-geometric theory it is that lies beyond $\N8$ supergravity 
and realizes these symmetries. Consequently, the rectification of charge 
mismatches above should not be viewed as a stop-gap kludge, but as an
important hint providing guidance towards  the right theory!

Agreement with the observed SM spectrum of spin-$\frac12$ fermions 
is  an encouraging step forward, but to make sure that this is not a mirage,
further evidence is needed that could potentially validate the proposal. 
And there is an option, at least in principle!  While a scheme based on $\N8$ supergravity
does not admit any new spin-$\frac12$ fermions, we are still left with 
eight massive gravitinos from (\ref{multiplet}). Because these carry
charges under SU(3)$\,\times\,$U(1) (and must also be subjected 
to a U(1) charge shift analogous to (\ref{cI})) it is clear that they must 
be extremely heavy on the one hand, and of extremely low abundance 
on the other hand, in order to have escaped detection until now.
While there is no chance of ever seeing such particles in collider experiments,
they can in principle be detected in underground experiments such as JUNO 
or DUNE (or possibly also in future dark matter search experiments).
A detailed study how to perform  such a search has been recently 
put forward in \cite{KLMN}. So at this point it is {\em wait and see}!

\vspace{0.3cm}

\noindent
{\bf Acknowledgments:} I would like to thank Anna Ceresole and Gianguido Dall'Agata 
for the invitation to contribute to this volume, and furthermore Gianguido Dall'Agata
for hospitality during a visit in Padova while finalizing this text.
I am very grateful to my collaborators in the adventurous story of maximal supergravity, 
in particular Bernard de Wit, Thibault Damour, Marc Henneaux, 
Axel Kleinschmidt, Krzysztof Meissner and Henning Samtleben, 
This work has been supported by the European Research Council (ERC) under the 
European Union's Horizon 2020 research and innovation programme (grant agreement No 740209).

\end{document}